# Towards Sustainable Energy Storage: Evaluating Polymer Electrolytes for Zinc Ion Batteries

Roya Rajabi, Shichen Sun, Booker Wu, Jamil Khan, Kevin Huang

Department of Mechanical Engineering, University of South Carolina, Columbia, South Carolina 29208, United States of America

**Abstract**

Polymer electrolytes present a promising solution to the challenges posed by aqueous electrolytes in energy storage systems, offering the flexibility needed for wearable electronics. Despite the increasing interest in polymer electrolyte-based zinc ion batteries (ZIBs), their development is still in its early stages due to various challenges. In this study, we fabricated three promising polymer electrolytes: CSAM (carboxyl methyl chitosan with acrylamide monomer), PAM (polyacrylamide monomer hydrogel electrolyte), and p-PBI (Phosphoric acid (PA)-doped polybenzimidazole) with $Zn(ClO_4)_2$ and $Zn(OTf)_2$, for their application in zinc ion batteries. Our results demonstrated that PAM hydrogel electrolyte exhibited very low LDH formation after a long cycle, demonstrating effective protection for zinc foil, and the high mechanical stability of the p-PBI membrane provided prolonged durability against short circuits through the formation of LDH. The presence of carboxyl groups in CSAM and the formation of O-H bonding facilitated ion movement, resulting in enhanced ionic conductivity, and preventing dendrite formation. Incorporating these hydrogels with high-performance zinc salts, such as zinc triflate ($Zn(OTf)_2$), resulted in impressive stability, with the symmetric cell demonstrating over 4000 hours of uniform and stable voltage profile under 1 mA/cm$^2$ and low overpotential of around 53 mV cycling with CSAM. The full-cell battery with PBI-T membrane showed the highest durability and capacity compared to CSAM-T and PAM-T, due to the greater availability of free protons for storing zinc in the cathode.

**Introduction**

In the pursuit of next-generation energy storage systems, aqueous zinc-ion batteries (AZIBs) stand out as a promising candidate featuring numerous advantages, for instance, the abundance and cost-effectiveness of zinc, high volumetric capacity (5851 mAh/cm$^3$), safety, conductivity of aqueous electrolyte (up to 150 mS/cm), as well as easy manufacturing [1, 2]. Despite these merits, zinc-ion batteries' practical commercialization still faces substantial challenges especially the performance degradation caused by the interaction between aqueous electrolyte and electrode. Specifically, the intrinsically abundant free water in aqueous electrolytes leads to inevitable side reactions towards the electrode, including hydrogen evolution reaction (HER), corrosion, and formation of side products, for example, zinc-layered double hydroxides (Zn-LDHs) [3, 4]. Moreover, the widespread use of glass fiber (GF) as a separator to retain the aqueous electrolyte introduces several issues. These include poor mechanical properties and the inevitable growth of zinc dendrites or Zn-LDHs due to the inhomogeneous pore distribution in GF. Consequently, these problems result in a shortened lifespan for aqueous zinc-ion batteries (AZIBs). [5].

Several methods have been investigated to address the aforementioned issues. The first route is the application of "water-in-salt" electrolytes (e.g. 30 m $ZnCl_2$) [6], hybrid electrolytes (e.g. $ZnSO_4$ and $MnSO_4$) [7], and organic electrolytes (e.g. zinc benzene sulfonamide $Zn(BBI)_2$) [8], delivering improved ionic conductivity and operation temperature range of aqueous electrolyte. The second route is replacing GF separators with functional separators [9] or polymer electrolytes [10] which provide better side reaction resistance and enhanced mechanical properties. Despite the advantages of both methods, certain drawbacks remain that obstacle the substitution of GF-aqueous electrolyte combination. For example, "water-in-salt" electrolyte has high costs and low mechanical durability. It carries the risk of precipitation and exhibits poor wettability on the anode surface [11]. Concentrated zinc salt enhances ionic conductivity and renders the system flexible across a wide temperature range due to the absence of strong H-bonding. In addition, The major problem of using organic electrolytes is the possible capacity loss due to the lack of $Zn^{2+}$ insertion. [12].

Among the above optimization methods, polymer electrolytes offer high thermal and mechanical properties suitable for harsh conditions while avoiding leakage issues with considerable cost-effectiveness. In addition, they showed a long life cycle due to the ability to immobilize anions, suppress dendrite growth, and increase cationic transport [13]. Various types of polymer electrolytes, including solid polymer electrolytes (SPEs) [14, 15], gel polymer electrolytes (GPEs) [16, 17], and hybrid polymer electrolytes (HPEs) [17, 18], are engineered for use in batteries. Solid polymer electrolytes poly(ethylene oxide) (PEO), offer high mechanical strength but often exhibit low ionic conductivity. In contrast, GPEs, such as xanthan, gelatine, polyacrylamide, and poly(vinylidene fluoride) (PVDF), are known for their high ionic conductivity but may lack sufficient mechanical stability [19]. Hybrid polymer electrolytes (HPEs) stand out as that combine the advantages of both SPEs and GPEs. By mixing two or more polymer electrolytes, such as carboxymethyl chitosan (CMC) and polyacrylamide (PAM) [20], it is possible to achieve a balance between ionic conductivity and mechanical strength, thus rendering enhanced the performance and reliability of batteries by optimizing the electrolyte composition [19].

The thickness and porosity are critical factors to enhance the ionic conductivity of polymer electrolytes. High porosity is essential for achieving high ionic conductivity. Additionally, attention must be paid to the Hofmeister effect of the anion in the chosen zinc salt and the compatibility of the polymer with the cathode [20]. This entails evaluating how chaotropic or kosmotropic anions exhibit different solvation effects, which stem from the salting-out effect of kosmotropes and the salting-in behavior of chaotropes [21]. Moreover, the thickness of the electrolytes plays a crucial role in determining the gravimetric energy density of the batteries [22].

Although the electrochemical performance of the polymer electrolytes has been improved a lot in recent years, seldom papers have focused on directly comparing the various polymer electrolytes under the same protocol. While individual studies emphasize the superior characteristics of specific polymer electrolytes, the absence of a consolidated and comparative analysis makes it challenging to discern the most effective and promising options for practical application. This highlights the need for a systematic investigation that assesses and compares the performance of different polymer electrolytes for ZIBs, providing a more nuanced understanding of their strengths, limitations, and overall suitability in diverse operational scenarios.

This paper thoroughly investigates and compare the effectiveness of three specific polymer electrolyte systems PAM, CSAM (Carboxymethyl chitosan with Polyacrylamide monomer), and p-PBI (Phosphoric acid (PA)-doped polybenzimidazole (PBI)), as the electrolyte of ZIBs. Each polymer electrolyte is meticulously examined for its electrochemical properties, mechanical properties, and overall performance. The performance of these polymer electrolytes was investigated with a high-performance zinc preinserted layered structure cathode $V_2O_5$. The results demonstrate that the PBI-T electrolyte exhibits the highest capacity compared to the CSAM-T and PAM-T electrolytes, with values of 325, 301, 278, 250, and 208 mAh/g at current densities of 0.3, 0.6, 1, 2, and 4 A/g, respectively. The PBI membrane also shows better stability, especially at low current densities (0.5 A/g). This work tends to provide a comprehensive study on a solid electrolyte, a gel electrolyte, and a hybrid electrolyte, which can be a helpful resource for practical application.

**Result and Discussion**

The structure of the PBI was demonstrated in Figure 1a, which is the combination of isophthalic acid and tetra aminobiphenyl (TAB). CSAM and PAM both have chitosan and acrylamide with different formulas and zinc insertion mechanisms. In CSAM hydrogel, the ratio of CMCS/AM (carboxymethyl chitosan to acrylamide monomer) is higher than in PAM hydrogel and a photo initiator was used for polymerization. Therefore, to introduce zinc ion as the charge carrier, the polymer was soaked in a zinc salt solution. In contrast, the ratio of CMCS/AM in PAM gel is low which caused a sticky structure and polymerized with a thermal initiator and zinc salt added as shown in Figures 1b and 1c. Among the common aqueous zinc salts applied in the field of AZIBs ($ZnNO_3$, $ZnCl_2$, $Zn(OTf)_2$, $ZnSO_4$, $Zn(ClO_4)_2$), zinc triflate, zinc sulfate, and zinc perchlorate, exhibited much better performance than the other zinc salt in aqueous electrolytes. Zinc perchlorate and zinc triflate showed high bulk ionic conductivity of 150 mS/cm and 65 mS/cm respectively [23]. The chaotropic anion in zinc perchlorate provides a better hydrogel improvement in mechanical properties, and zinc triflate has a near-neutral pH value (~6) which enables it to operate at a wider operating voltage. Consequently, zinc perchlorate and zinc triflate have been chosen in this work. As reported in Ref [23] 2 M zinc perchlorate and 1 M zinc triflate have demonstrated the highest ionic conductivity in contrast to other concentrations. Hence, these concentrations have been employed to incorporate the zinc salts into the polymer electrolytes. In this study, polymers incorporating zinc perchlorate and zinc triflate have been denoted as CSAM-P, PAM-P, PBI-P, and CSAM-T, PAM-T, and PBI-T, respectively.

The image and morphology of the polymer electrolytes are presented in Figure 1d-1j. Figures 1d,1e, and 1f exhibit the images of CSAM after being soaked in 2 M zinc perchlorate for 24 hours, the PBI, and PAM hydrogel, respectively. As reported in Ref [20] a ternary interaction occurs among the chaotropic $ClO_4^-$ anions, water, and CSAM hydrogel. These visuals offer a comparative insight into how each polymer interacts or absorbs the zinc perchlorate solution, providing a visual understanding of their behaviors and structural changes upon exposure to the solution. Evident from the observation is the expansion of CSAM during immersion, a compelling demonstration of the formation of ternary bonds between CSAM, water, and $ClO^{4-}$ ions. While it was not observed when soaked in zinc triflate.

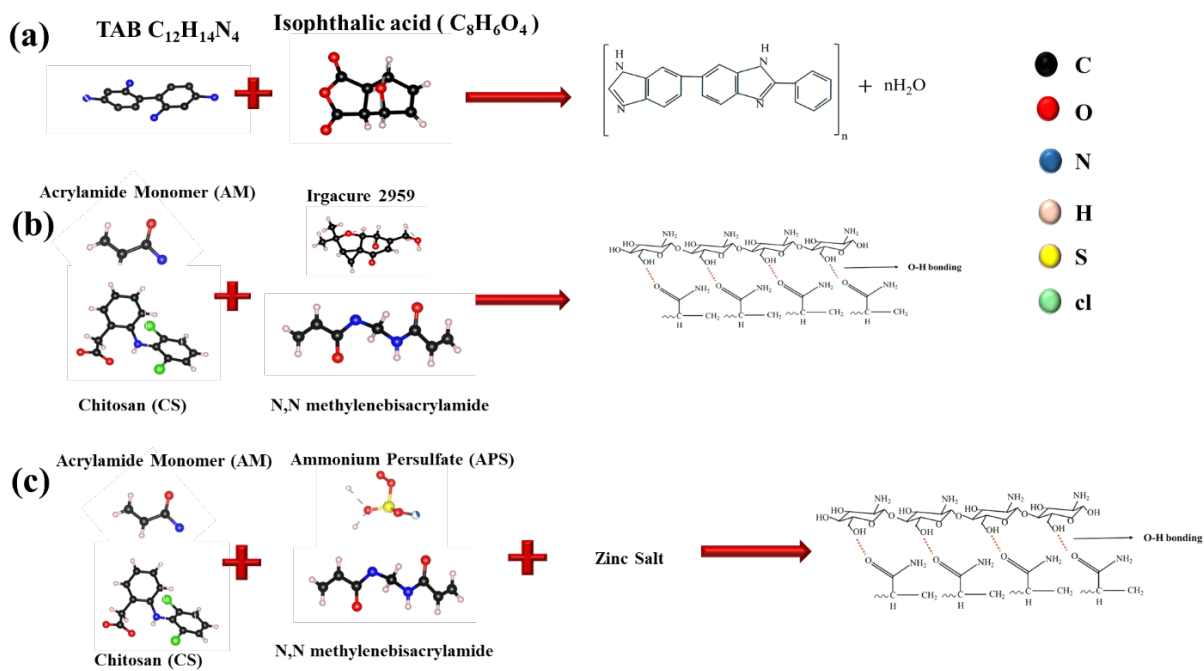

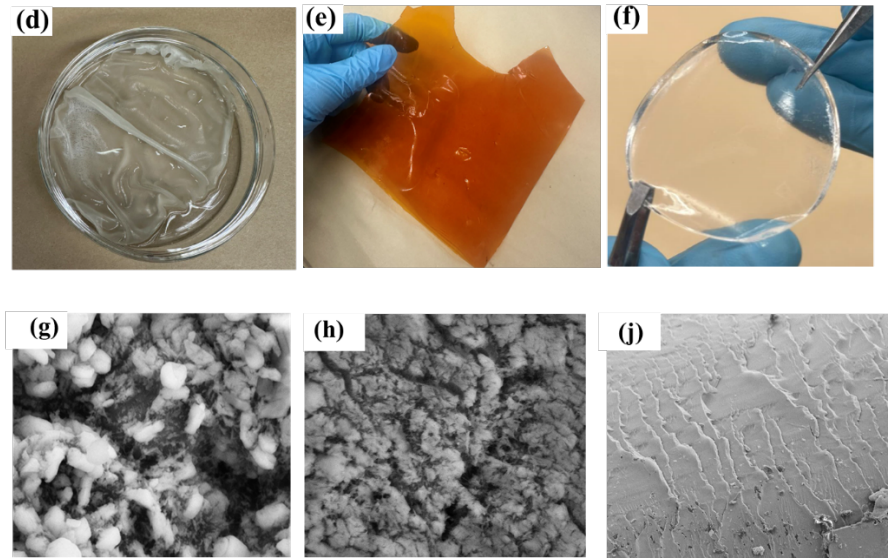

Figure 1: Schematic diagram of polymers' structural components and dynamic interactions of a) p-PBI, b) CSAM, c) PAM; Photo Image of d)CSAM, e)PBI, f)PAM polymer electrolyte; SEM image of g)CSAM, h) PBI and i)PAM polymers.

The composition of CSAM is mainly carboxymethyl chitosan and the introduction of acrylamide monomer markedly boosts the polymer's ionic conductivity and mechanical properties, primarily augmenting its crucial characteristics of porosity and elasticity [24]. These microstructural distinctions are vividly depicted in Figure 1 g-j. Quantitative analysis of the images using ImageJ software revealed porosity percentages of 28% for CSAM, and 26% for PBI, and due to denseness, no network is observable in PAM. It accounts for the fact that showcasing the material's variations in structural features among different polymer compositions. Moreover, the hydroxyl group within chitosan actively interacts with water and the anion of the zinc salt. This interaction forms ternary bonding, reducing free water availability and minimizing hydrogen bonding. Consequently, this configuration boosts the material's resilience to lower temperatures, enhancing its overall tolerance.

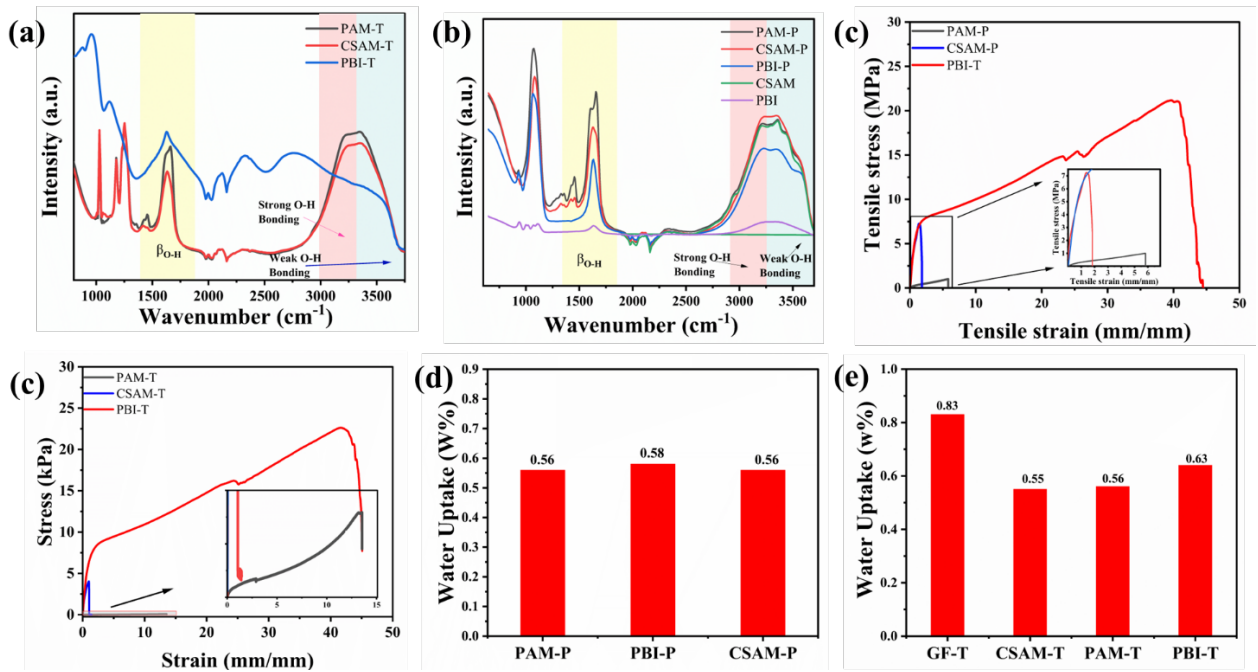

*Figure 2: a-b) FTIR spectra of polymer membrane and polymer membrane with zinc perchlorate and zinc triflate, c-d) Strain-stress curve of CSAM, PBI, PAM with two zinc salts, e-f) Effect of polymer structure and zinc salt on water uptake.*

To assess the distinct impact of two zinc salts (2 M zinc perchlorate and 1 M zinc triflate) on three polymers, an FTIR test was employed. In CSAM and PAM hydrogel, water within the structure prompts a distinct O-H bonding peak (3000 to 3600 cm$^{-1}$), absent in the PBI polymer membrane. Upon soaking CSAM in zinc perchlorate (as depicted in Figure 2a), a notable increase in the intensity of the weak O-H bonding peak (3300 to 3600 cm$^{-1}$) is evident. These interactions play a crucial role in limiting free water content, thus restraining dendrite formation during battery cycling. In the case of PBI, the observed peaks stem from the diffusion of the zinc salt solution into its pores, mirroring similar spectral patterns. Interestingly, the effect of zinc triflate differed among CSAM, PAM hydrogels, and the PBI membrane. Figure 2a illustrates the detection of a distinctive peak at the wavenumber 1255 cm$^{-1}$, indicating C-F stretching in the sample. Soaked samples in zinc triflate lack the weak O-H bonding peak for all polymers which demonstrates the better ternary O-H bonding.

PBI stands out as a highly promising membrane in fuel cell technology owing to its exceptional mechanical and thermal stability [25]. A comparative analysis of this polymer with two hydrogels highlighted its distinctive characteristics, showcasing an impressive high tensile strain (39%) and stress (21 MPa), as depicted in Figure 2c,d. Notably, no discernible impact of zinc salt on the mechanical properties of PBI was observed. In contrast, CSAM and PAM exhibited enhanced mechanical stability in the presence of zinc perchlorate, displaying increased elongation and tensile stress due to ternary O-H formed. The modulus elasticity of the polymers was calculated based on the formula presented in the supporting document. Particularly, both CSAM-P (17.5 MPa) and PAM-P (0.3157 MPa) demonstrated superior modulus of elasticity compared to those of CSAM-T (9.5 MPa) and PAM-T (0.023 MPa). The modulus of elasticity for PBI was calculated to be 5.61 MPa, further underlining its robust mechanical properties within this comparison.

Extended soaking times tend to enhance the ionic conductivity of polymers, but they can adversely impact their mechanical stability [20]. Therefore, in this study, we opted for an optimized soaking duration of 24 hours for CSAM and PBI membranes. With this immersion duration established, we delved into comparing the water/liquid content among these polymers. The batteries equipped with liquid electrolytes and glass fiber separators displayed a higher liquid content (> 80%) than the polymers. This heightened liquid presence signifies freer water, potentially leading to corrosion on the zinc foil surface.

Notably, both PBI and CSAM exhibited similar liquid content (58% w and 56%w ) as shown in Figures 2e and 2f, for both zinc salt, while for PAM, the water content in the formula (56%w) was considered as the liquid content of the polymer. This comparative analysis reveals the varying degrees of liquid presence within these materials, highlighting the potential impact on their performance, especially concerning corrosion on zinc foil surfaces caused by water. [26].

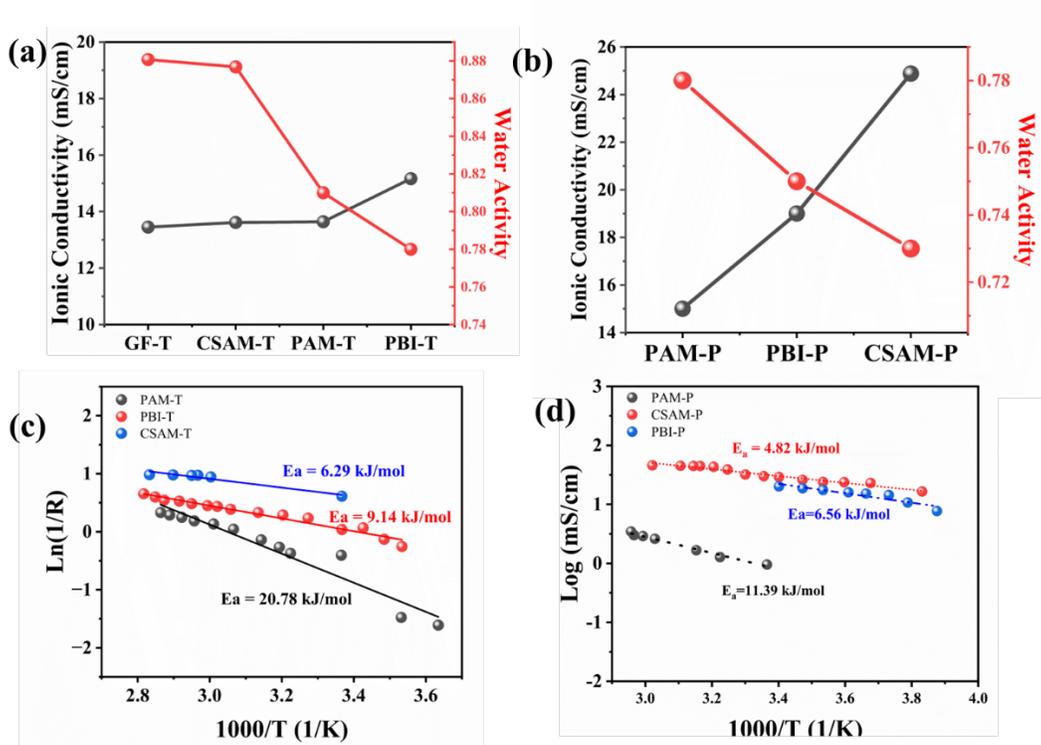

Figure 3: Ionic Conductivity and water activity of the polymer electrolytes a) Ionic conductivity and water activity of the CSAM-T, PBI-T, GF-T, b) Ionic conductivity and water activity of CSAM-P, PBI-P, PAM-P c) Ionic conductivity of CSAM-P, PBI-P, PAM-P from -5 to 60 °C, d) PAM-T, PBI-T, CSAM-T from -5 to 60 °C.

Regarding the electrochemical properties and ionic conductivity of polymer electrolytes, both the solid-like mechanism of the polymer and the liquid-like mechanism with ion mobility are crucial characteristics [13]. For instance, 2 M zinc perchlorate and 1 M zinc triflate exhibit high bulk conductivities of 60 and 150 mS/cm, respectively (Ref[23]). However, when incorporated into a cell with a glass fiber separator, 1 M zinc triflate demonstrated a significantly lower conductivity of around 19 mS/cm, as shown in Figure 3a.

Figures 3a and 3b illustrate the comparison of ionic conductivity and water activity among three polymer electrolytes PBI, CSAM, and PAM. CSAM showed better conductivity and less water activity. With zinc triflate, CSAM, PBI, PAM, and glass fiber exhibit ionic conductivities around 15, 14, 13.5, and 13 mS/cm respectively (Figure 3a), while with zinc perchlorate, these values shift to 24, 19, and 15 mS/cm respectively (Figure 3b). The difference in water activity among the polymers helps explain why CSAM might have better ionic conductivity than others. CSAM's carboxyl group can retain water, reducing accessible water and preventing corrosion. This trapped water helps ions move more easily, lowering the overall water activity in CSAM. Moreover, the ionic conductivity of the PBI membrane is partially attributed to its high hydrogen conductivity (7.5 mS/cm) and the storage of the electrolyte in its porous structure. while the PAM hydrogel electrolyte exhibits the lowest ionic conductivity.

The polymers' ionic conductivity was measured across temperatures from -5 to 70 °C. CSAM-T, PBI-T, and PAM-T exhibited activation energies of 6.29 kJ/mol, 9.14 kJ/mol, and 20.78 kJ/mol, respectively. When paired with zinc perchlorate, these values changed to 4.82 kJ/mol, 6.56 kJ/mol, and 11.39 kJ/mol in the same order as shown in Figures 3c and 3d. PAM exhibited the highest activation energy compared to that of the PBI and CSAM, no matter what zinc salt was applied. More importantly, the data illustrates that zinc perchlorate resulted in lower activation energy for dissolving $Zn^{2+}$, and the CSAM-T is the lowest among all samples. The lower activation energy of CSAM-P suggests better stability across temperatures, as anticipated because of the formation of ternary hydrogen bonds and the breaking of water's O-H bonds in CSAM-P. This means this CSAM-P electrolyte can function well over a broader range of temperatures.

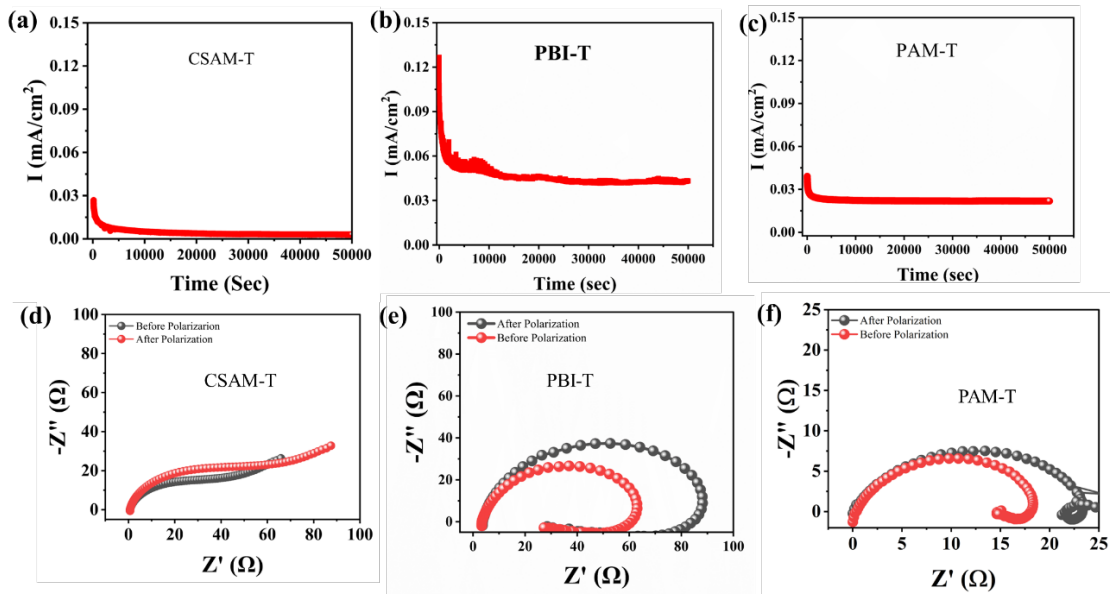

*Figure 4: ) Zn$^{2+}$ transference number characterization of the a,d) CSAM-T, b,e) PBI-T, c,f) PAM-T polymer electrolyte with the applied voltage of 10 mV.*

The dual-ion conductivity exhibited by these electrolytes underscores the importance of understanding the conductivity of zinc ions. Consequently, we conducted measurements to determine the transference number of zinc ions in selected polymers. The symmetrical cells Zn cells were employed to evaluate the Zn$^{2+}$ transference number and stripping/plating performance of PBI-T, PAM-T, and CSAM-T electrolytes. The transference number was calculated based on the Vincent–Evans equation presented in the supporting document. As calculated in Figures 4a-f the battery with PBI-T, PAM-T, and CSAM-T electrolytes can provide Zn$^{2+}$ transference numbers of 0.27, 0.52, 0.64. Higher transference numbers in CSAM and PAM hydrogel electrolytes attributed to the carboxylate groups in the gel can provide special channels for ion transport to ensure sufficiently high ionic conductivity [27].

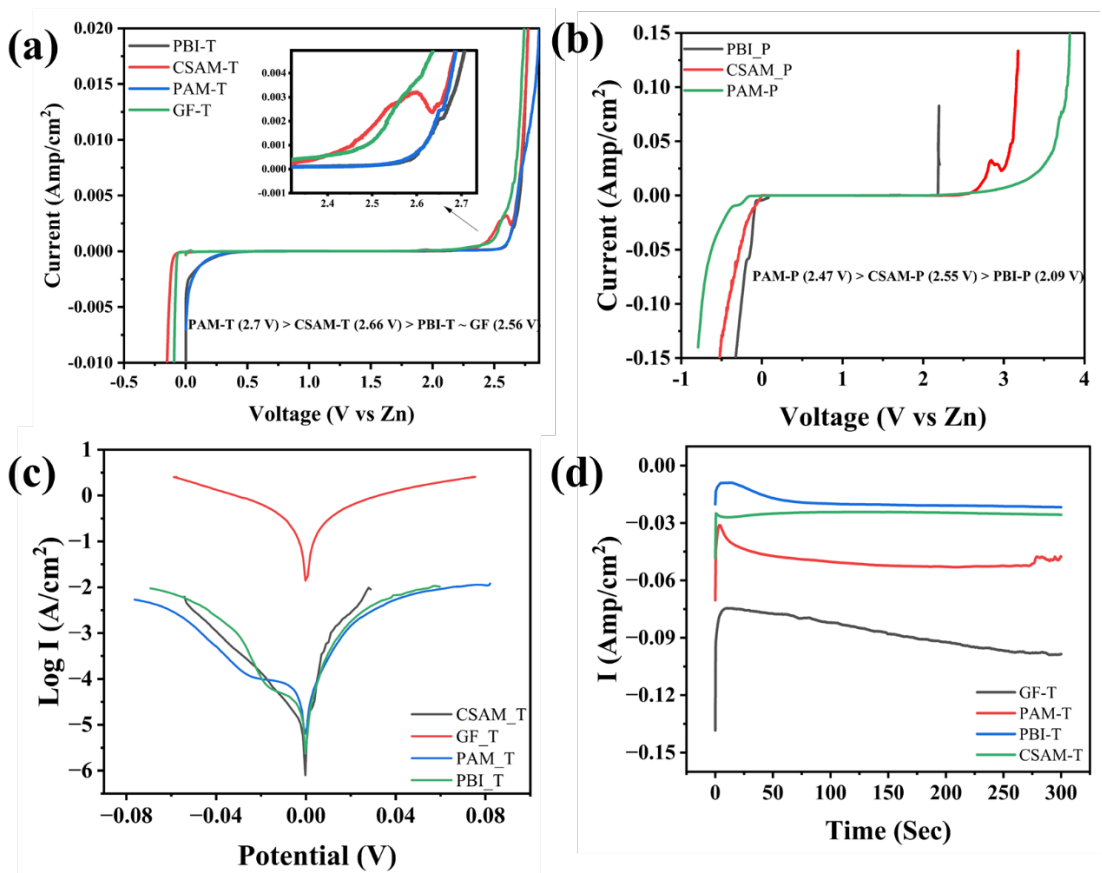

*Figure 5: a-b) Stability potential window of three polymer electrolytes with zinc perchlorate and zinc triflate c) Comparison of linear polarization curves presenting the corrosion for bare Zn with different polymer electrolyte and liquid electrolyte with glass fiber d) the voltage-time curves during Zn nucleation at -200 mA cm$^{-2}$ in Zn/GF-T/Zn, Zn/CSAM-T/Zn, Zn/PAM-T/Zn, Zn/PBI-T/Zn*

With lower water activity and less accessible water, batteries tend to have a stability potential window compared to aqueous electrolytes. In our experiments, PAM displayed the widest stability potential windows among the polymer electrolytes due to its lower water absorption compared to other polymers and glass fiber. Polymers paired with zinc triflate showed a broader operating voltage range due to their higher pH value compared to those with zinc perchlorate [23]. A nearly neutral pH avoids starting the oxygen evolution reaction (OER) in acidic electrolytes and the hydrogen evolution reaction (HER) in alkaline ones. Retarded (OER) and (HER) of the polymer electrolytes compared to liquid electrolytes is an indication of the corrosion resistance ability. This fact is confirmed further by conducting the three-electrode experiment conducted as described in supporting information. As shown in Figure 5c, using zinc triflate with a wider operating voltage, the corrosion current of the liquid electrolyte with glass fiber as the separator is much higher than polymer electrolyte and the PAM polymer electrolyte has the lowest corrosion current. Regarding deposition behavior, chronoamperometry was applied to further study the nucleation mechanism in which the movement of $Zn^{2+}$ ions can be observed through I-t (current-time) graphs, conducted under an overpotential of -200 mV (seen in Figure 5b). In symmetric cells with aqueous electrolytes, the fluctuating I after 300 seconds indicates erratic or 3D diffusion, leading to excessive dendrite growth. Conversely, the consistent current (I) value suggests a 2D diffusion process on zinc foil for CSAM-T, PBI-T, and PAM-T due to a more uniform porous structure compared to glass fiber. CSAM-T showed a constant current after 12 seconds, PAM-T reached a constant value after 20 seconds, and PBI-T showed a constant current after one minute, which demonstrated a better zinc deposition using CSAM-T hydrogel compared to PAM-T and CSAM-T. This consistent $Zn^{2+}$ diffusion is visually explained through the $Zn^{2+}$ ion pathways in Figure 5d. Among polymer electrolytes, CSAM-T demonstrated a strong affinity for polar $Zn^{2+}$ and a higher bonding strength with the Zn atom. The electrostatic attraction facilitates easy ion movement, guiding the continuous transport of $Zn^{2+}$ ions. Consequently, this results in uniform nucleation and deposition on the Zn anode's surface. Higher initial stripping and plating potentials of Zn is a demonstration of faster kinetics in electrolytes that contain water [28]. PBI polymer showed a higher initial potential than CSAM and PAM hydrogel and all of them had higher potential than liquid electrolyte with glass fiber.

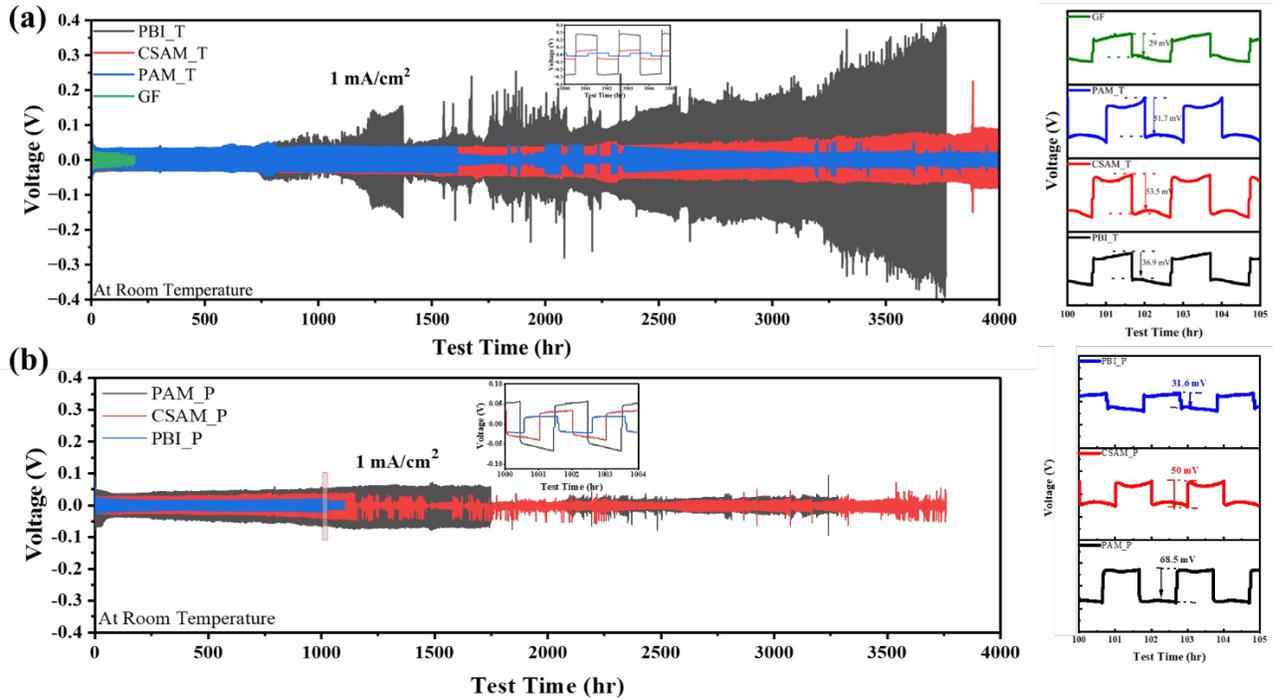

*Figure 6: Cycling performance of the symmetrical Zn cells with CSAM, PAM, and PBI polymer electrolyte under 1 mA/cm² and at room temperature a) with zinc triflate b) with zinc perchlorate*

The interfacial stability between the electrolyte and Zn metal anodes is evaluated by symmetric Zn-Zn cells at room temperature under a current density of 1 mA/cm². The voltage profiles of Zn plating/stripping are displayed in Figure 6. This experiment is a good illustration of the electrochemical and diffusion characteristics of the polymer electrolytes under investigation. The employment of glass fiber as a separator in the cell revealed limited durability. Figure 6a illustrates the superior performance of polymers containing zinc triflate when compared to those coupled with zinc perchlorate. Notably, all three polymers with zinc perchlorate exhibited a consistent stripping/plating behavior within the initial 1200 hours with an overpotential of 31.6, 50, and 65.5 mV for PBI-P, CSAM-P, and PAM-P respectively. However, batteries utilizing polymer electrolytes with zinc triflate demonstrated a longer durability, CSAM-T presents very stable cycle performance, showing uniform Zn deposition without change in overpotential (around 53.5 mV) for more than 4000 hours. PAM-T had a similar performance with some trivial short circuits and an overpotential of 51.5 mV. However, PBI continues to work, obvious overpotential hysteresis is observed in the cell after approximately 1000 h, indicating the instability of the interface layer and its heavy accumulation on the Zn foil surface.

The regular overpotential and voltage hysteresis is possibly attributed to the propensity of layered double hydroxide (LDH) dissolution in batteries employing PAM and CSAM electrolytes. The symmetrical battery, constructed with a PBI membrane, demonstrated prolonged cycling without experiencing a shortage, albeit exhibiting a notable increase in overpotential starting from ~37 mV to 800 mV. This phenomenon can be attributed to nonuniform Zn deposition and the formation of a substantial layer of $Zn_x(OTf)_y(OH)_{2x-y}\cdot nH_2O$, forming a rough surface on electrodes while using zinc triflate, resulting in an elevation of electrical resistance. To prove this hypothesis the XRD pattern and SEM image of the anodes after cycles have been examined as shown in Figures 7 and 8. The accrual of this layer is deemed responsible for the observed augmentation in overpotential within the battery system.

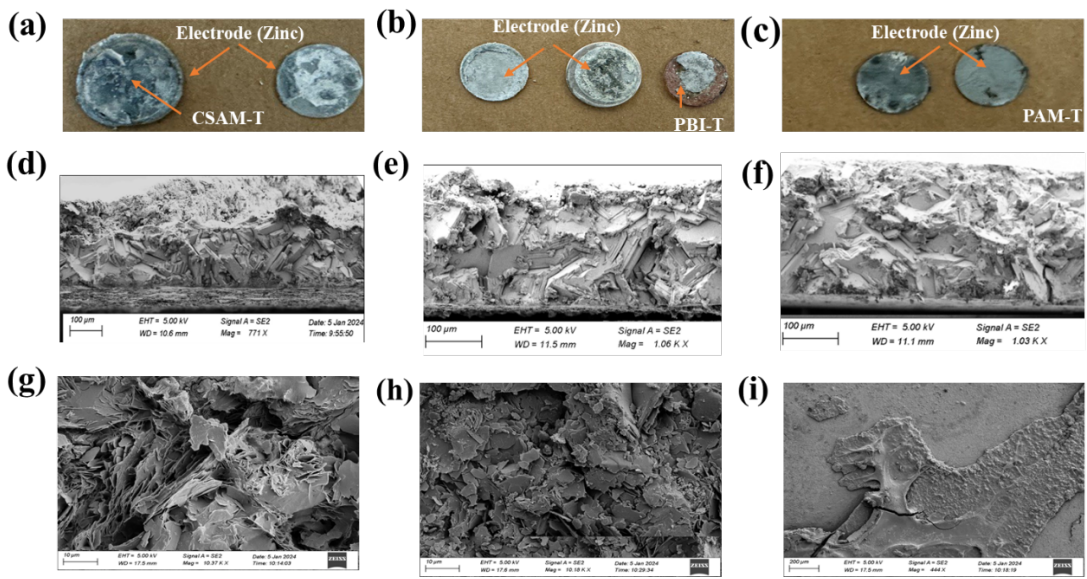

*Figure 7: Image of the electrodes after cycling a) Electrodes cycled with CSAM-T b) Electrodes cycled with PBI-T c) Electrodes cycled with PAM-T. d-l) SEM image of cross-section and surface of electrodes after cycling with CSAM-T, PBI-T, and PAM-T.*

In pursuit of a comprehensive understanding of the diverse performance exhibited by these polymer electrolytes in a half-cell battery, a post-test analysis has been undertaken. From this section onward, we omitted the investigation of zinc perchlorate and focused on zinc triflate, as both would ultimately lead to the same conclusion.

A tough surface was observed on electrodes cycled with CSAM and PBI electrolytes as shown in Figures 7a and 7b, though PBI electrolytes remained non-damaged after 4000 hours of cycling while a thick layer of flakes attached to the anode assembled with PBI electrolyte. However, as illustrated in Figure 7c smooth surface was observed on the zinc foil employed as an anode in conjunction with a PAM electrolyte. On the other hand, CSAM was delaminated which agrees with the mechanical property discussion as mentioned in Figure 2 to show the PBI's better mechanical strength. The microstructure examination of the electrodes' cross-section and surface in Figures 7d-i shows the usage of a portion of the zinc. Surface analysis of the electrodes (Figures 7g-i) demonstrates more flakes on the electrode when cycled with CSAM-T and PBI-T due to having more water (proton) accessible with zinc foil and the formation of zinc hydroxide and zinc salt hydroxide.

To further understand the nature of the formed Zn-LDHs under the influence of electrical current, we performed post-test analysis on the electrically cycled samples. The XRD and XPS patterns in Figure 8 and Figure 9 suggest that the Zn-anode surfaces are covered with Zn-LDHs, the same as the by-products in the aqueous electrolyte as previously reported [23].

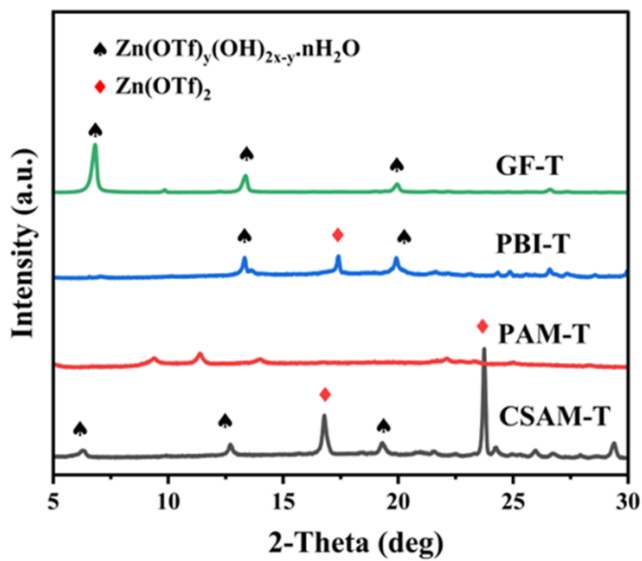

*Figure 8: XRD pattern of zinc foil after cycling with PBI-T, CSAM-T, PAM-T, and zinc triflate as liquid electrolyte.*

The PBI membrane lacks carboxyl bonding, it does not exhibit any affiliation with aqueous electrolytes, and functions as a separator. In this context, its robust mechanical properties, along with tolerance against tip diffusion and short-circuit inhibition, contribute to prolonged cycle stability. A uniform porous structure compared to glass fiber, facilitating even deposition, and providing tolerance against short circuits, proves beneficial for achieving extended durability. It is noteworthy that PBI can retain less water compared to glass fiber, thereby aiding in slower LDH formation. Additional advantages of using PBI membrane such as commendable mechanical and thermal stability, enabling the battery to operate for extended durations, even in high pressure and higher or lower temperature than the normal conditions if it is used with an appropriate zinc salt. Importantly, PBI's inability to trap aqueous electrolytes within its polymer structure leads to faster LDH formation when compared to hydrogel electrolytes like CSAM and PAM. To gain deeper insights into the distinct stability of polymer electrolytes, pure zinc, and cycled zinc, X-ray photoelectron spectroscopy (XPS) was employed. The analysis focused on three polymer electrolytes. As illustrated in Figure 9, the examination revealed that the uncycled zinc and the anode cycled with PAM-T exhibited a low amount (~1.6%) of zinc in the form of zinc oxide. In contrast, anodes cycled with CSAM-T and PBI-T contained higher proportions of zinc oxide (10.6% and 8.1%, respectively), along with zinc hydroxide.

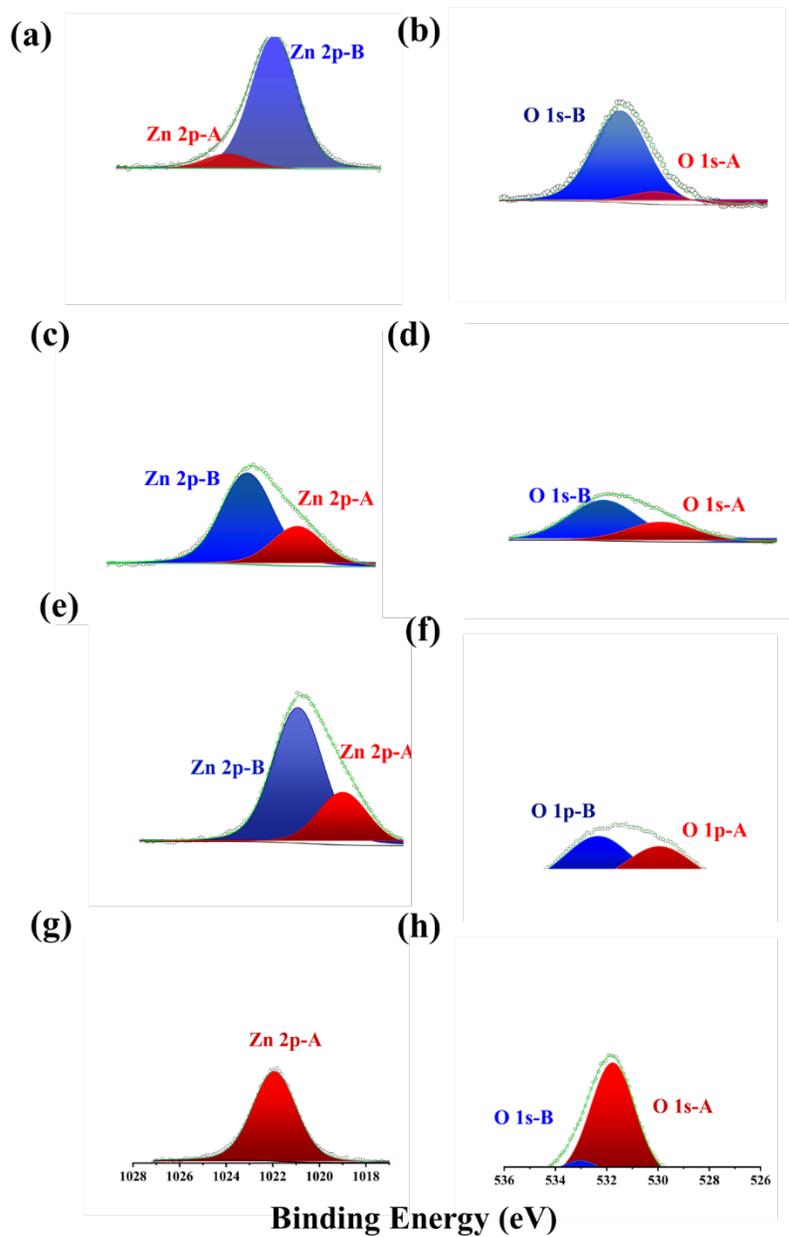

*Figure 9: a) O1s of pure zinc b) Zn 2p XPS spectrum of pure zinc c) O1s of zinc anode cycled with CSAM-T d) Zn 2p XPS spectrum of zinc anode cycled with CSAM-T, e) O1s of zinc anode cycled with PBI-T f) Zn 2p XPS spectrum of zinc anode cycled with PBI-T, g) O1s of zinc anode cycled with PAM-T h) Zn 2p XPS spectrum of zinc anode cycled with PAM-T*

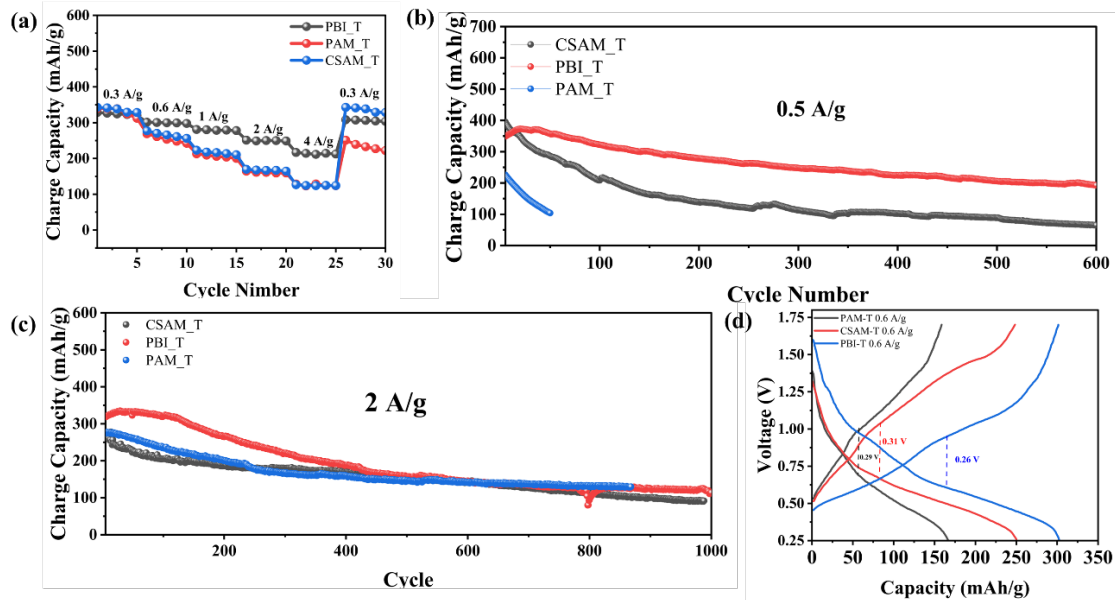

*Figure 10: Practical applications of the full zinc-ion batteries with CSAM-T, PBI-T, PAM-T electrolyte at room temperature. a) Cycling performance of the Zn/ZnVO battery at different current density at room temperature. b) Rate performances of Zn/ZnVO battery c-d) Cycling performance of the Zn/ZnVO battery with CSAM-T, PBI-T, PAM-T electrolyte under 0.5 A g$^{-1}$ and 2 A g$^{-1}$*

To analyze the behavior of polymer electrolytes in a full-cell battery, $Zn_{0.1}V_2O_5 \cdot nH_2O$ (ZnVO) xerogel cathode was utilized. The cathode was synthesized following the method outlined in Ref [29]. Within this cathode, the $Zn^{2+}$ charge transfer process occurs within a shrinking gallery, concomitant with the dissociation of gallery water into $H^+$ and $OH^-$ ions. Consequently, a dual $Zn^{2+}/H^+$ storage mechanism within the ZnVO with a bilayer structure has been proposed [29]. These findings underscore the significance of proton involvement in the charge transfer process.

The performances of the Zn//ZnVO batteries with CSAM-T, PAM-T hydrogel electrolytes, and PBI-T separator were detected at room temperature. The results show that the rate performances of the battery with PBI-T at different current densities of 0.3, 0.6, 1, 2, and 4 A/g are then provided, with the stable capacity of 343, 289, 283, and 249 mA h g$^{-1}$ respectively and for CSAM-T and

PAM-T it is around 324, 268, 220, 170, 127 mAh/g. After 25 cycles when the rate goes back to the lowest (0.3 A/g) battery with PBI-T, CSAM-T, and PAM-T membrane electrolytes showed the capacity of 339, 310, and 234 mAh/g as shown in Figure 10a which shows better reversibility of cathode with PBI-T separator. One possible reason can be due to the high proton conductivity of the PBI membrane which provides the essential component of the storage mechanism. Long-term stability test results under 0.5 and 2 A/g current are demonstrated in Figures 10b and 10c. With the rate of 2 A/g, batteries with PBI-T, CSAM-T, and PAM-T showed the initial capacity of 319, 276, and 260 mAh/g and they were able to cycle 1000 times with a capacity of higher 100 mAh/g. At a current density of 0.5 A/g, PBI-T and CSAM-T were able to cycle 600 times with initial capacities of 353, and 390 mAh/g while keeping its capacity of more than 100 mAh/g, with a better performance of battery with PBI-T electrolyte; however, the battery equipped with PAM-T at low current density of 0.5 A/g, and initial capacity of 219 mAh/g was not able to cycle more than 50 times due to the dense structure of the PAM and not having enough proton for the storage mechanism. The voltage difference of the charge and discharge curve of these polymer electrolytes at a specific current density (0.6 A/g) and cycle number were compared. Lower voltage difference demonstrates better reversibility. Battery with PBI electrolyte shows the lowest voltage difference of 0.26 V, followed by 0.29 and 0.31 V for CSAM-T and PAM-T electrolyte as shown in Figure 10d. The results show that PBI-T membrane has more reversibility and capacity than CSAM-T and PAM-T electrolytes. The reason stems from its charge transfer mechanism. PBI-T's zinc conductivity comes from the storage of electrolytes in its porous structure; therefore, it can provide more accessible water which is beneficial when coupled with a cathode with dual $Zn^{2+}/H^+$ storage mechanisms.

For further investigation into the behavior of full-cell batteries utilizing PBI-T and CSAM-T, the surfaces of the anode and cathode were scrutinized following 1000 cycles under 2 A/g. XRD examination of the anode reveals the formation of LDH in the PBI-T-cycled anode, while no LDH is detected in the CSAM-T-cycled anode, as illustrated in Figure 11a. Furthermore, the XRD pattern of the cathode cycled with PBI-T (Figure 11b) differed from that of pure ZnVO as depicted in Figure S4, indicating the insertion of flakes into the gallery spacing and underscoring battery degradation. It can be concluded that the reason for the degradation of the battery for PBI membrane is the LDH formation. The faster degradation of CSAM and PAM electrolytes is attributed to the lack of protons, leading to a rapid capacity decrease before LDH formation and subsequent battery depletion. In contrast, in the PBI membrane, LDH formation can interfere with zinc storage reactions. While the low $Zn^{2+}$ transference number of the PBI membrane provides enough protons for the storage mechanism, it also contributes to LDH formation, potentially impacting battery performance.

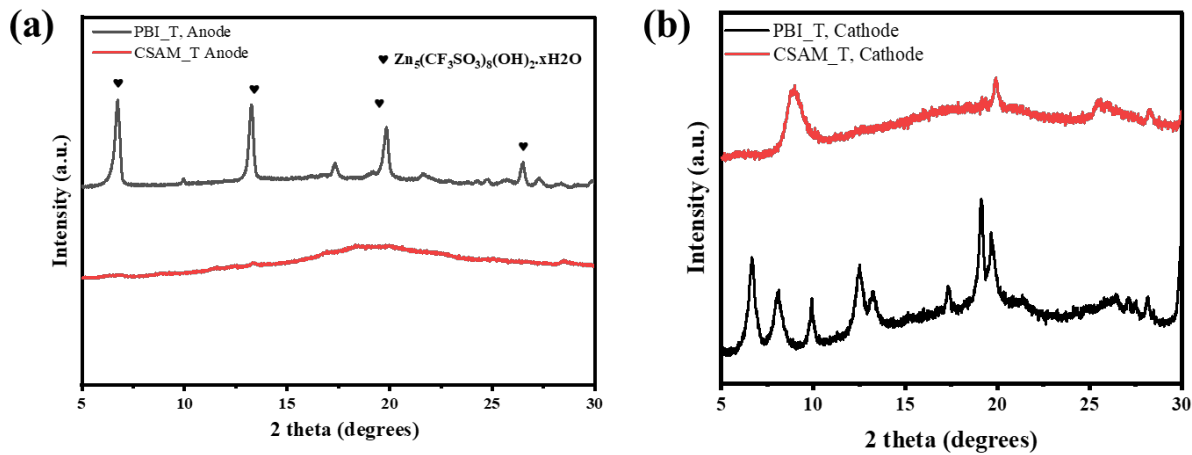

Figure 11: XRD pattern of the a) zinc anode and b) cathode after 1000 cycles under 2 A/g with PBI-T and CSAM-T

A brief comparison between this work and some other papers worked on solid/semi solid electrolyte represented in Table S2. The electrolytes used in this work (PBI-T, CSAM-T, PAM-T) exhibit high ionic conductivities (13.6-15 mS/cm), which are competitive with or superior to many other studies. Notably, they outperform electrolytes like PHP880 (0.02 mS/cm) and PVHF/Mxene-g-PMA (0.269 mS/cm) but are lower than PSAZn-gel (59 mS/cm). Comparing the initial capacities achieved in this work (283-340 mAh/g at 2 A/g) relative to other works with similar rate are significantly higher than many other cathode materials.

The stability of this cathode with the three polymer electrolyte at low current density which is more challenging (0.5 A/g) is significantly longer than other reports and the long durability presented are at high current density mostly.

**Conclusion**

Three polymer electrolytes include CSAM and PAM hydrogel and PBI were compared electrochemically to find the advantages and disadvantages of each electrolyte for flexible and wearable zinc-ion batteries. PAM and CSAM hydrogel electrolytes regulate the amount of water that effectively inhibits side reactions and dendrite formation. PBI membrane as a strong membrane with a high stability has a benzene structure therefore does not show chemical reaction with zinc salt solution and porous structure of the membrane holds the electrolyte and it can be categorized as a separator. Higher pH value and stability potential windows of zinc triflate caused a better performance in the battery especially with CSAM which caused more than 4000 hours of stability in a symmetric cell. Due to having zinc salt in the PAM structure and not using the soaking method for zinc insertion the polymer solution after polymerization and having a dry surface PAM showed a low ionic conductivity while in a symmetric cell it shows the lowest amount of LDH formation because of having a dry surface in contact with zinc foil (anode). In full cell battery testing, with $Zn_{0.1}V_2O_5$, CSAM and PBI showed good stability and high capacity in a full battery testing using $Zn_{0.1}V_2O_5$ under current density of 2 A/g with initial capacity of over 300 mAh/g while keeping the capacity over 100 mAh/g for 1000 cycles. At low current density of 0.5 A/g PBI showed higher efficiency and better reversibility which stems from providing more $H^+$ which helps a better zinc storage. In this research we achieved a long stable electrolyte for more than 4000 hours with CSAM and PAM with zinc triflate as a zinc salt with a uniform stripping and plating voltage profile. PBI membrane as a separator showed a good possible option for zinc ion batteries due to high mechanical and thermal stability while due to not having an anti-dendrite feature it needs a protective layer on the zinc anode to avoid LDH formation. Hydrogel showed a good anti-dendrite feature while water based structure caused a weaker structure compared to PBI membrane.

# Towards Sustainable Energy Storage: Evaluating Polymer Electrolytes for Zinc Ion Batteries

## 1. Experimental Section

*Material:*

Zinc perchlorate hexahydrate ($Zn(ClO_4)_2 \cdot 6H_2O$) and zinc trifluoromethyl sulfonate ($Zn(CF_3SO_3)_2$), both obtained from Sigma-Aldrich (98%), served as the primary chemical sources in this study. Carboxymethyl chitosan (CMCS), acrylamide (AM), 2-hydroxy-4'-(2-hydroxyethoxy)-2-methylpropiophenone (Irgacure 2959), N,N'-methylenebisacrylamide (MBAA), and ammonium persulfate were also procured from Sigma-Aldrich Co. Furthermore, zinc foil was sourced from Qingyuan Metal Co., Ltd. The synthesis of ZnVO involved ammonium metavanadate ($NH_4VO_3$), sodium dodecyl sulfate (SDS), polyvinylidene fluoride (PVDF), and super carbon, while sulfuric acid was obtained from Sigma-Aldrich Co.

### 1.1. Synthesis of Polymer Electrolytes

*Synthesis of CSAM:*

To create the CSAM hydrogel electrolyte, 0.888 grams of carboxymethyl chitosan (CMCS) were dissolved in 7 milliliters of water at room temperature. Following this, 1.777 grams of acrylamide (AM), 2959.2 milligrams of 2-hydroxy-4'-(2-hydroxyethoxy)-2methylpropiophenone (Irgacure) as an initiator, and 1.8 milligrams of N,N'-methylenebisacrylamide (MBAA) were added to the CMCS solution. The resulting mixture was gently stirred until a uniform and homogeneous precursor solution was achieved. To eliminate any trapped air bubbles, the solution underwent centrifugation for 5 minutes at a speed of 1200 rotations per minute (rpm). After this preparation step, the solution was cast onto a glass sheet using a doctor blade. Subsequently, the solution underwent photo-initiation to facilitate the formation of the CSAM hydrogel. Finally, the CSAM-P, and CSAM-T hydrogel electrolytes were obtained by immersing it in a 2M zinc perchlorate and 1M zinc triflate solution for a duration of 24 hours.

*Synthesis of PBI:*

The process commenced with the preparation of a PA-doped p-PBI membrane through the PPA sol-gel process [30], facilitating the attainment of a high PA doping level in the p-PBI. Following this, the membrane underwent multiple soakings in deionized water to completely eliminate polyphosphoric acid. To transform the p-PBI membrane into a Zn-ion conductor, the resulting p-PBI membrane was immersed in a 1 M $Zn(CF_3SO_3)_2$ solution and 2 M $Zn(ClO_4)_2$ for 24 hours to maximize electrolyte impregnation.

*Synthesis of PAM:*

5 ml of water and 25 mg of CMCS should be stirred until a homogenized solution is obtained. The next ingredients to be added are 1.5 g of AM, 20 mg of ammonium persulfate, 1 mg of MBA, and 3.72 grams or 3.6 grams of zinc triflate or perchloride. Mix the blend until a uniform solution is achieved. Transfer the mixture onto a glass plate and leave it there for two hours at 60 degrees Celsius.

**1.2. Zinc foil preparation.**

In this study, full and half coin cell tests were conducted using Zn discs (Sigma-Aldrich, >99.9%, 250 μm thickness) in 13 mm. All Zn samples were polished using 1200 grid sandpaper and sonicated to remove any materials on the surface before testing.

## 1.3. Synthesis of Cathode Materials

**Synthesis of $Zn_{0.1}V_2O_5 \cdot nH_2O$ (ZnVO) Xerogel:** The synthesis of the ZnVO xerogel was accomplished using a hydrothermal method. In a standard procedure, 5 mmol of ammonium metavanadate ($NH_4VO_3$) was dissolved in 40 mL of deionized water, initially stirred slowly, and subsequently subjected to vigorous stirring at 90 °C for 2 hours. Simultaneously, 5 mmol of sodium dodecyl sulfate (SDS) were dispersed in 10 mL of deionized water. Upon achieving a transparent solution, 25 mL of a 1.5 M zinc sulfate solution were added, resulting in a 35 mL solution. The two solutions were thoroughly mixed, and the pH was adjusted to approximately 2.5 by adding 1 M sulfuric acid.

Following stirring at room temperature for 20 hours, the resultant dark orange solution was transferred into an autoclave equipped with a Teflon liner and maintained at 175 °C for 10 hours. Subsequent to the reaction, the dark-green precipitate was collected, washed meticulously with deionized water and ethanol 8 times, and centrifuged every time for 5 mins under 1200 rpm, and ultimately dried at 70 °C for 12 hours. The detailed synthesis procedure is illustrated in Figures S1 and S2 for reference. This method ensures the controlled formation of the ZnVO xerogel, offering a systematic approach for its reproducible preparation.

**Cathode Preparation:** The initiation of the cathode preparation involved the mechanical grinding of 120 milligrams of ZnVO and 52 milligrams of carbon for a duration of thirty minutes. This composite mixture was then integrated with 1 milliliter of polyvinylidene fluoride (PVDF) and N-methyl-2-pyrrolidone (NMP) solution, prepared at a concentration of 0.03 grams per milliliter, following vigorous mixing for a period of two hours. After three days of continuous stirring to ensure homogeneity, the cathode material was cast onto a pre-cut and pressed stainless-steel mesh. Subsequently, the cathode assembly underwent a drying process in a vacuum oven at a controlled temperature of 90 °C overnight. Post-drying, the cathode was accurately weighed and subjected to a pressing step as an integral part of the final preparation procedures, as depicted in Figure S3. This systematic approach guarantees the consistent and reproducible fabrication of the cathode material for subsequent utilization in battery configurations.

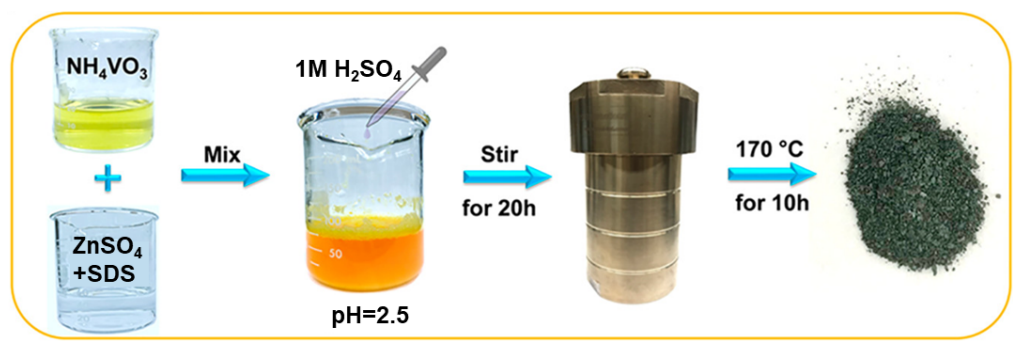

Figure S12: Schematic illustration of preparation of the ZnVO powder

*Figure S1: Schematic illustration of preparation of the ZnVO powder.*

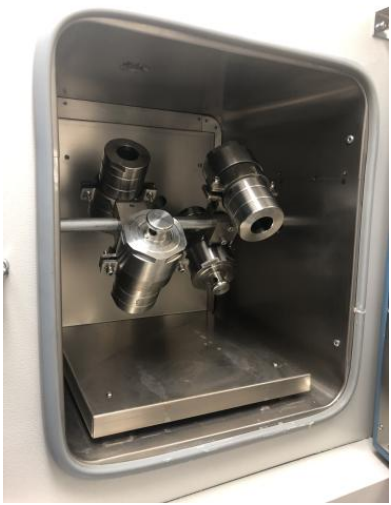

*Figure S2: Schematic of the rotating oven used for the reaction step of the ZnVO preparation.*

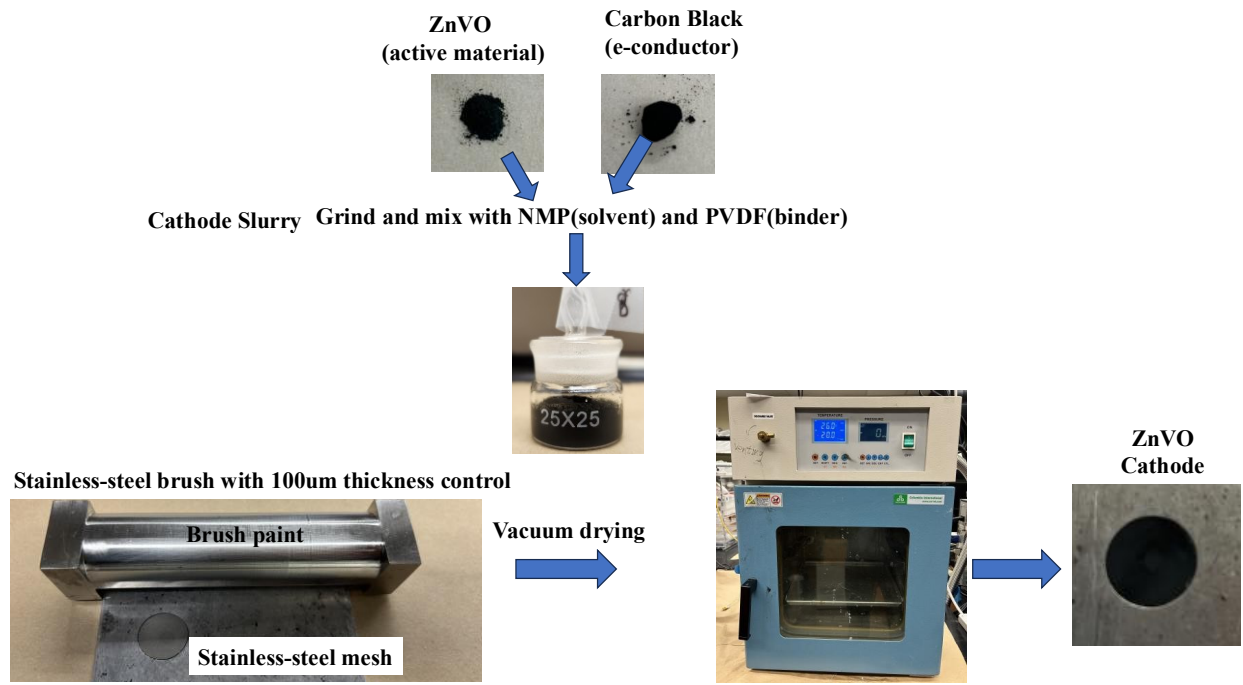

*Figure S3: Schematic preparation of the cathode production.*

**Characterization of the ZnVO powder:**

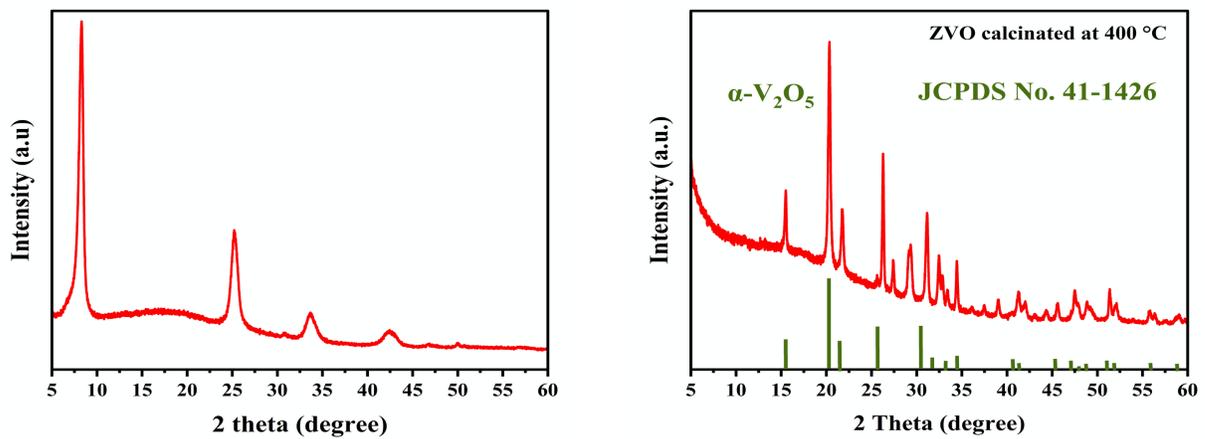

*Figure S4: Characterization of the ZnVO powder a) XRD pattern of ZnVO electrode b) The XRD patterns of ZnVO powders calcinated at 400°C.*

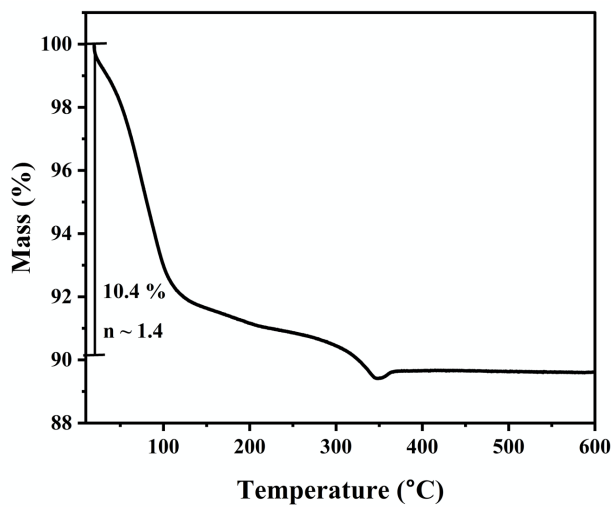

*Figure S5: TGA curves of ZnVO under air atmosphere at a heating rate of 2 C/min.*

## 2. Material Characterization

***Water uptake measurement:*** The ability of a hydrogel electrolyte membrane to absorb water is represented by its water uptake ratio (WU). A higher water (thus Zn salt solution) update enhances the membrane's ionic conductivity. To determine this property, the samples were first weighted in the wet state and then dried in an oven until a constant weight was achieved. The water uptake (WU) is calculated by:

$$WU\% = \frac{W_w - W_d}{W_w} 100 \quad (1)$$

where $W_w$ and $W_d$ are the weights in wet and dry states, respectively.

***Fourier-transform infrared spectroscopy (FTIR)***: It is well known that $Zn^{2+}$ forms solvation shell with $H_2O$ molecules, which plays a role in Zn/electrolyte interactions. To assess the solvation strength of the hydrogel electrolytes, we employed FTIR (Agilent, Cary-630) to measure the stretching vibration of O–H bonding as an indicator of the solvation strength. We focus on the wavenumber range of 3200 to 3600 cm$^{-1}$ in FTIR spectrum, which is referred to as the "OH stretching region" and contains information on hydrogen bonding.

*Water Activity:* The water activity of the hydrogel electrolytes was measured by AQUALAB 4TE water activity meter at ambient temperature.

*Mechanical Test:* The mechanical properties of the hydrogel electrolyte were evaluated using a testing machine (UTM2103, Shenzhen Suns Technology Co., Ltd.) at a rate of 30 mm/min at room temperature. Young's modulus which describes the relationship between stress (force per unit area) and strain (proportional deformation in an object), can be obtained using below formula:

$$E = \frac{\sigma}{\epsilon} = \frac{stress}{strain} \tag{2}$$

*Surface characterizations:* X-ray diffraction (XRD, RigakuD/MAX-2100) and field emission scanning electron microscope (FESEM, Zeiss Gemini-500) and X-ray photoelectron spectroscopy (XPS) measurements conducted for surface analysis. XRD patterns were collected with Cu Kα radiation (λ = 1.5418 Å) from 5° to 35° with an interval of 0.02° and a scan speed of 5° min$^{-1}$. FESEM were acquired for the morphologies and compositions of the products formed on the surfaces of Zn foils during soaking and electrochemical measurements. X-ray photoelectron spectroscopy (XPS) measurements were performed on XPS research facility at University of South Carolina.

### 3. Electrochemical characterization

*Ionic Conductivity:* To measure the ionic conductivity of the hydrogel electrolyte, we first pouched the hydrogel membrane 0.35 mm, then sandwiched it with two stainless steel disks. The conductivity was measured by an AC impedance method in a frequency range from 100 kHz to 1 Hz using Solartron 1260/1287 electrochemical workstation.

***Hydrogel Transference Number***: Zn-ion transport number reveals the percentage of total ionic conduction attributed to Zn-ion. Electrolytes with higher transference numbers can mitigate concentration polarization and dendrite growth. Moreover, a higher transference number of fosters more efficient charge transfer between the electrolyte's cathode and anode, leading to improved overall efficiency. Transference number can be calculated using steady state current approach in a symmetric cell. In a high-conductivity polymer battery, it is preferable for the zinc transference number to be near to one. The transference number ($t_+$) can be computed as follows:

$$t_+ = \frac{I_s}{I_0} \qquad (3)$$

where $I_o$ and $I_s$ represent the initial and steady-state cell currents, respectively, to consider the effect of oxidation on the electrodes that can alter the experiment results, Vincent–Evans equation is used that uses a correction factor to characterize the system before and after polarization, and the equation was rewritten as:

$$t_+ = \frac{I_s(\Delta V - R_0 I_0)}{I_0(\Delta V - R_s I_s)} \qquad (4)$$

$\Delta V$ is the polarization voltage of 10 mV, $R_0$ and $R_s$ are the bulk resistance of electrolytes before and after the polarization, respectively. $I_o$ ascertain the initial current, and $I_s$ is the steady-state resistance current of the sample, the electrolyte film was analyzed via EIS both before and after DC polarization [31, 32].

***Electrochemical Window Stability:*** The electrochemical stability window of the polymer electrolytes was assessed through cyclic voltammetry (CV), employing stainless steel electrodes and zinc foil. The determination of the potential range associated with the oxygen evolution reaction (OER) and hydrogen evolution reaction (HER) involved subjecting the cells to both positive (0 to 3 V) and negative (0 to -3 V) voltage sweeps. This methodological approach facilitates a systematic investigation of the electrolyte's capacity to withstand varying voltage regimes, providing crucial insights into its electrochemical stability under different conditions.

***Three-electrode cell configuration:*** This test was adopted for electrochemical measurements; figure S6 shows the schematic of the setup. A circular Zn foil (diameter: 13mm) was used as the working electrode (WE), and a platinum foil (Sigma-Aldrich, 99%), in parallel to the WE, was used as the counter electrode (CE), while a standard Ag/AgCl (Pine research) was used as the reference electrode (RE). Electrochemical measurements including potentiodynamics and chronoamperometry (-200 mV) were carried out using a Solartron 1260/1287 electrochemical workstation. The open circuit voltage (OCPs) was first recorded, followed by potentiodynamics and EIS. For potentiodynamics measurement, the potential was swept at a rate of 10 mV s$^{-1}$ within ± 0.15 V vs OCP to avoid side reactions (e.g. hydrogen evolution or oxygen evolution reactions or chlorine oxidation reaction) at either WE or CE.

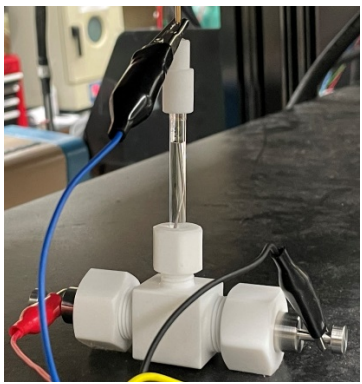

*Figure S6: Schematic illustration of three-electrode setup.*

***Symmetrical Cell Testing:*** To evaluate the stability of Zn/hydrogel electrolytes, symmetrical cells of CR2032 type coins were assembled and tested under galvanostatic charging and discharging cycles at ±1 mA cm$^{-2}$ with a duration of 1 hour per cycle using a LAND battery testing system (CT2001A). The cells were allowed to rest for 24 hours before testing, and a 3-minute rest was given after each cycle. The termination voltage of the cycling was set to be ±2.1 V.

***Full Battey Testing***: To assess the stability of a full cell battery (Zn/hydrogel/ZnVO), the cell was assembled in a configuration resembling symmetrical cells of CR2032 type coins. The battery was subjected to galvanostatic charging and discharging cycles using a LAND battery testing system (CT2001A) at various currents including 0.3, 0.6, 1, 2, 4, and 6 A/g. Each cycle had a duration of 1 hour and was performed with the following parameters. Prior to testing, the cells were allowed to rest for 24 hours. Additionally, a 3-minute rest period was provided after each cycle. The cycling process was terminated when the voltage reached a range between 0.25 and 1.5 V.

*Table 1: Comparison of this work with some other polymer electrolytes developed for zinc ion battery*

| Ref | Electrolyte | Ionic Conductivity [mS/cm] | Cathode | Initial Capacity [mAh/g] | Stability |
|---|---|---|---|---|---|
| This Work | PBI/ Zn(OTf)$_2$ | 15 | Zn$_{0.1}$V$_2$O$_5$.nH$_2$O | 340 (2 A/g) | 1000 (~70%) |
|  | CSAM/ Zn(OTf)$_2$ | 13.6 |  | 289 (2 A/g) | 1000 (~70%) |
|  | PAM/ Zn(OTf)$_2$ | 13.6 |  | 283 (2 A/g) | 1000 (~70%) |
| [33] | PHP880/ Zn(OTf)$_2$ | 0.02 | γ-MnO$_2$ | 65 (0.5C) | 395 (87.6%) |
| [34] | PVHF/Mxene-g-PMA/Zn(OTf)$_2$ | 0.269 | MnHCF | 90 (2C) | 10000 (80%) |
| [34] | IL-PAM/ Zn(ClO$_4$)$_2$ | 15 | LFP | 110 (10C) | 1000 (80%) |
| [35] | SA-PAM/ ZnSO$_4$ | 29.2 | Na$_{0.65}$Mn$_2$O$_4$.1.3H$_2$O | 120 (2 A/g) | 1000 (97%) |
| [36] | P(ICZn-AAm)/ ZnSO$_4$ | 2.15 | V$_2$O$_5$ | 271 (2C) | 150 (~100%) |
| [37] | ZIG/ Zn(TFSI)$_2$ | 2.6 | MnHCF | 100 (1C) | 200 (~100%) |
| [38] | Polymer Glue Zn-PG/ ZnSO$_4$ | 6.18 | NH$_4$V$_4$O$_{10}$ | 150 (5C) | 2000 (~100%) |
| [37] | EMIMBF$_4$ | 16.9 | V$_6$O$_{13}$ | 360 (0.4A/g) | 200 (80%) |
| [38] | Polymer Glue Zn-PG/ Zn(OTf)$_2$ | 6.18 | NH$_4$V$_4$O$_{10}$ | 150 (5C) | 2000 (~100%) |
| [39] | ZCEs/ Zn(TFSI)$_2$ | 0.059 | V$_2$O$_5$ | 134.7 | 80 (99%) |
| [40] | PSAZn-gel/ Zn(TFSI)$_2$ | 59 | NH$_4$V$_4$O$_{10}$ | 300 (0.5 A/g) | 200 (97%) |
| [41] | (PAX-G)/ZnSO$_4$ | 16.8 | V$_2$O$_5$ | 117 (0.117 A/g) | 500 (60%) |
| [21] | PAM/ Zn(ClO$_4$)$_2$ | 24.9 | Cu$_x$V$_2$O$_5$·nH$_2$O | 150 (10 A/g) | 1000 (~100%) |

| [42] | CSAM/ Zn(ClO$_4$)$_2$ | 19 | PANI | 80 (5 A/g) | 2000 (~100%) |
| [43] | PBI/ Zn(OTf)$_2$ | 2.2 | V$_2$O$_5$ | 385 (3 A/g) | 2000 (92%) |

*Table 2: Summary of comparison in electrochemical properties of three polymer electrolytes*

|  | Ionic Conductivity [mS/cm] | Activation Energy [kJ/mol] | Mechanical Stability [MPa] | Water Activity | Stability Potential Windows | Electrolyte Uptake | Transference Number | Symmetric cell Stability [hr] | Initial Capacity [mAh/g, @ 2A/g] | Stability in full cell battery testing [hr] |
|---|---|---|---|---|---|---|---|---|---|---|
| CSAM-T | 15.1 | 6.29 | 9.5 | 0.78 | 2.66 | 0.56 | 0.62 | > 4000 | ~260 | 1000 |
| PBI-T | 13.6 | 9.14 | 21 | 0.8 | 2.56 | 0.65 |  | ~1000 | 321 | 1000 |
| PAM-T | 13.6 | 20.78 | 1 | 0.82 | 2.7 | 0.56 |  | >1500 | ~260 | ~890 |
| GF-T | 13.4 | -- | -- | 0.88 | 2.56 | 0.83 |  | < 200 | -- | -- |